\documentclass[12pt]{article}
\usepackage{setspace}
\usepackage{fullpage}
\usepackage{graphicx}
\usepackage{subfigure}
\usepackage{bm}
\usepackage{amssymb,amsmath}
\usepackage{setspace}
\usepackage{float}
\usepackage[super,sort&compress]{natbib}

\usepackage{color}
\definecolor{darkblue}{rgb}{0,0,0.4}
\usepackage[colorlinks=true,citecolor=black,linkcolor=darkblue,
urlcolor=darkblue,bookmarks=false,bookmarksopen=false,
pdfpagemode=None,pdfstartview=FitH]{hyperref}

\begin{document}

\newcommand{\URL}[1]{\href{#1}{#1}} 

\renewcommand\refname{References}

\title{{\Large MoS$_2$ as an ideal material for valleytronics:
valley-selective circular dichroism and valley Hall effect}}

\maketitle
\begin{center}
Ting Cao,$^1$ Ji Feng,$^{1\dagger}$
Junren Shi,$^1$ Qian Niu$^2$  \& Enge
Wang$^{1*}$ 
\par\end{center}

{\begin{center}
$^1$International Center for Quantum Materials, Peking
University, Beijing, China 100871. $^2$Department
of Physics, University of Texas, Austin, TX 178712, USA. $^{\dagger}$E-mail:
\texttt{jfeng11@pku.edu.cn.}\textcolor{black}{{} $^{*}$E-mail:}\texttt{ egwang@pku.edu.cn}
\par\end{center}}

\pagebreak{}
\centerline{ABSTRACT}

A two-dimensional honeycomb lattice harbors a pair of inequivalent
valleys in the\textbf{ k}-space electronic structure, in the vicinities
of the vertices of a hexagonal Brillouin zone, \textbf{K}$_{\pm}$.
It is particularly appealing to exploit this emergent degree of freedom
of charge carriers, in what is termed ``valleytronics'',\cite{Gunawan06,Rycerz07,Akhmerov07}
if charge carrier imbalance between the valleys can be achieved. The
physics of valley polarization will make possible electronic devices
such as valley filter and valley valve, and optoelectronic Hall devices,\cite{Xiao07,Yao08}
all very promising for next-generation electronic and optoelectronic
applications. The key challenge lies with achieving valley imbalance,
of which a convincing demonstration in a two-dimensional honeycomb
structure remains evasive, while there are only a handful of examples
for other materials.\cite{Takashina06,Bishop07,Eng07,Zhu11} We show
here, using first principles calculations, that monolayer MoS$_2$,
a novel two-dimensional semiconductor with a 1.8 eV direct band gap,\cite{Mak10,Splendiani10}
is an ideal material for valleytronics by valley-selective circular
dichroism\cite{Yao08}, with ensuing valley polarization and valley
Hall effect.

\pagebreak{}

A non-equilibrium charge carrier imbalance between valleys is the
key to creating valleytronic devices. The principal mechanism invoked
here is circularly polarized optical excitation.\cite{Xiao07,Yao08}
In this approach, the two valleys absorb left- and right-handed photons
differently, a phenomenon referred to as circular dichroism (CD).
In order to obtain the valley-selective CD, it is essential then to
break the inversion symmetry of the honeycomb lattice. In the case
of graphene, it was suggested that by interacting graphene with a
substrate such that the center of inversion can be obliterated, whereupon
a gap opens up in each valley.\cite{Giovannetti07,Zhou07} This strategy,
however, is quite challenging experimentally. And even in the eventual
realization, there can only be a weak perturbation to graphene via
the weak covalent coupling at large van der Waals separations.\cite{Zhou07} 

Since its first isolation, monolayer MoS$_2$ has attracted
immense attention. Many measurements have been performed to characterize
the optical and transport properties of this material.\cite{Mak10,Splendiani10,Lee10,RamakrishnaMatte10,Radisavljevic11}
In monolayer MoS$_2$, two layers of sulfur atoms in
a two-dimensional hexagonal lattice are stacked over each other in
an eclipsed fashion. Each Mo sits in the center of a trigonal prismatic
cage formed by 6 sulfur atoms (\textcolor{blue}{Fig.1a}). Quite remarkably
in the context of current discussion, the natural stable structure
of free-standing monolayer MoS$_2$ is a honeycomb lattice
with inequivalent bipartite coloring, breaking the inversion symmetry
(see \textcolor{blue}{Fig.1b}). Bulk MoS$_2$ has an
indirect band gap. Interestingly, when thinned to the monolayer limit
the material acquires direct band gaps located exactly at the corners
of the Brillouin zone. Indeed, with its 1.8 eV direct band gap and
unique two-dimensional structure embracing the high-symmetry valleys,
monolayer MoS$_2$ has much to offer in the exploration
of novel electronic and optielectronic devices and the associated
physics. It is quite natural to question whether it is possible to
achieve valley-selective CD in this semiconducting atomic membrane,
which will endow electrons in this material the valley degree of freedom,
in addition to charge and spin that have been routinely explored in
conventional device physics.

\begin{spacing}{1}
\begin{center}
\begin{figure}[H]
\centering{}\includegraphics[scale=0.36]{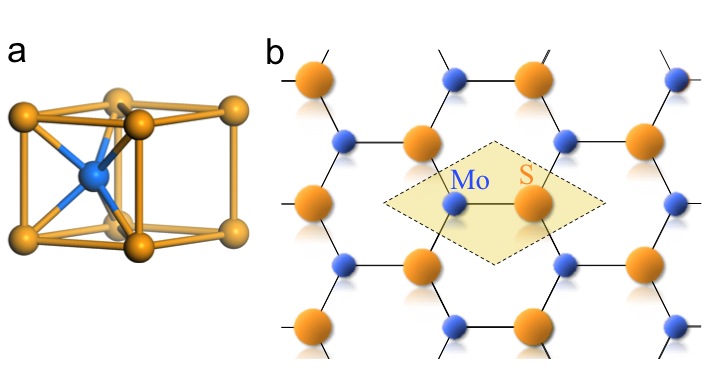}\label{fig:structure}
\caption{
{\small The crystal structure of monolayer MoS$_2$.
(a) Coordination environment of Mo in the structure. (b) A top view
of the lattice, emphasizing the connection to a honeycomb lattice.
In our calculations, we use an optimized structure at the level of
local density approximation in density functional theory (see text).
The shaded region bounded by dashed lines corresponds to one primitive
cell. The unit cell parameter is }\textit{\small a}{\small{} = 3.12
\AA, and the vertical separation between sulfur layers is 3.11 \AA.
}}
\end{figure}

\end{center}
\end{spacing}

The key quantity to assess is, therefore, the ${\bf k}$-revolved
degree of optical polarization, $\eta({\bf k})$, between the top
of the valence bands and the bottom of the conduction bands\cite{Yao08}

\begin{equation}
\eta({\bf k},\omega_{cv})=\frac{|\mathcal{P}_{+}^{cv}({\bf k})|^{2}-|\mathcal{P}_{-}^{cv}({\bf k})|^{2}}{|\mathcal{P}_{+}^{cv}({\bf k})|^{2}+|\mathcal{P}_{-}^{cv}({\bf k})|^{2}}.
\end{equation}
This quantity is the difference between the absorption of left- and
right-handed lights ($\pm$), normalized by total absorption, at each
${\bf k}$-point and evaluated between the top of the valence bands
($v$) and the bottom of conduction bands ($c$). Note that the dependence
on the transition energy, $\hbar\omega_{cv}({\bf k})=\epsilon_{c}({\bf k})-\epsilon_{v}({\bf k})$,
is implicit through ${\bf k}$. Here, the transition matrix element
of circular polarization is $\mathcal{P}_{\pm}^{cv}({\bf k})=\frac{1}{\sqrt{2}}[P_{x}^{cv}({\bf k})\pm iP_{y}^{cv}({\bf k})]$.
The interband matrix elements, ${\bf P}^{cv}({\bf k})=\langle\psi_{c{\bf k}}|\hat{{\bf p}}|\psi_{v{\bf k}}\rangle$,
are evaluated using the density functional perturbation theory (DFPT),
within the local density approximation (LDA),\cite{Ceperley80} as
implemented in {\small VASP}. \cite{Gajdos06} Spin-orbit coupling
is not included in our calculations. Briefly, a planewave basis set
is employed at a cut-off energy 600 eV, and a total of 80 bands are
included to ensure convergence of all computed quantities. A very
dense ${\bf k}$-point mesh ($123\times123$ grid points) over the
reducible two-dimensional full Brillouin zone is sampled in our calculations.

As shown in \textcolor{blue}{Fig.2a} chiral absorption selectivity
is indeed exact at ${\bf K}_{\pm}$ with $\eta=\pm1$. The contrast
in the chiral absorptivities between the valleys owes its origin in
the symmetry of both lattice and local atomic orbitals, which is quite
different from the gapped graphene case\cite{Yao08}. A band state
relevant to the optical excitation originates from local atomic states
(or more generally, Wannier functions) bearing different orbital magnetic
quantum numbers, $l$. In this case, the states at the top of the
valence bands involve only $d_{x^{2}-y^{2}}$ and $d_{xy}$ on Mo,
and $p_{x}$ and $p_{y}$ states on S. At ${\bf K}_{+}$, the $d$-states
on Mo hybridize as $\frac{1}{\sqrt{2}}(d_{x^{2}-y^{2}}+id_{xy})$
($l=+2$) to interact with $\frac{1}{\sqrt{2}}(p_{x}+ip_{y})$ ($l=+1$).
At ${\bf K}_{-}$, the $d$-states on Mo hybridize as $\frac{1}{\sqrt{2}}(d_{x^{2}-y^{2}}-id_{xy})$
($l=-2$) to interact with $\frac{1}{\sqrt{2}}(p_{x}-ip_{y})$ ($l=-1$).
At the bottom of the conduction band, only $d_{z^{2}}$ state on Mo
($l=0$) is involved. These atomic orbitals form such linear combinations
in accordance with the $D_{3h}$ point group symmetry and the lattice
translational symmetry. 

\begin{spacing}{1}
\begin{center}
\begin{figure}[H]
\centering{}\includegraphics[scale=0.33]{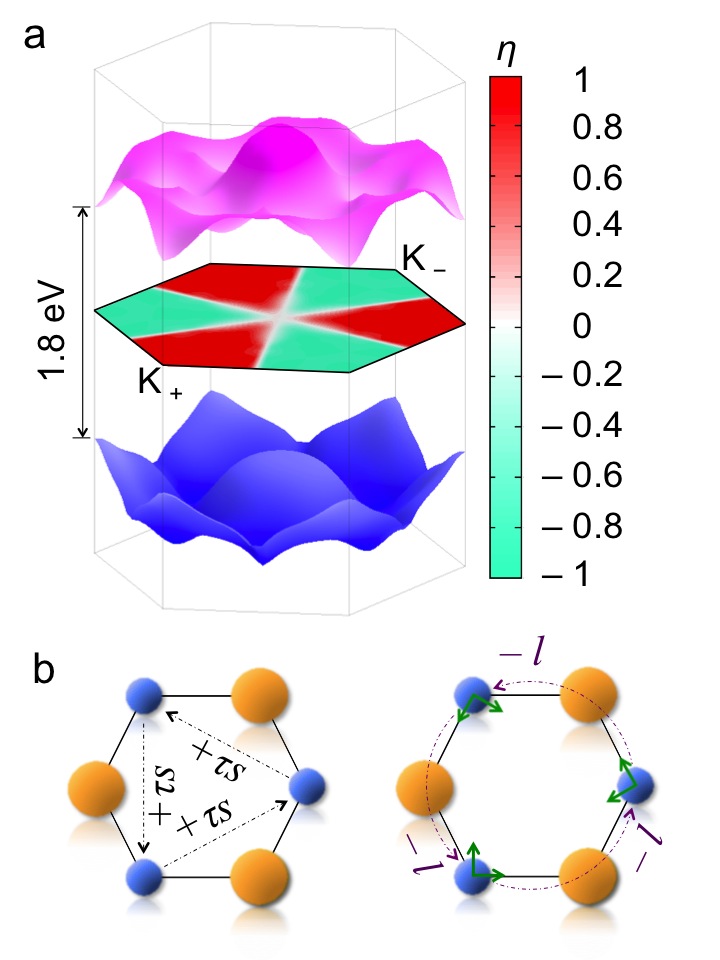}\label{fig:rotor}\caption{{\small Valley-selective circular dichroism of monolayer MoS$_2$.
(a) Top valence band (blue) and bottom conduction band (pink). The
center hexagon is the Brillouin zone color-coded by the degree of
circular polarization, $\eta$(k), as defined in the text. The vector
connecting }\textbf{\small K}{\small $_{+}$ and }\textbf{\small K}{\small $_{-}$
is perpendicular to Mo-S bond in the crystal structure in (a). (b)
Schematic of phase winding on the MoS$_2$ lattice
that gives rise to the chiral optical selectivity. Left panel: the
contribution to phase winding from the Bloch lattice phase, where
$\tau=\pm1$ is the valley index, and $s=1,$ 2 corresponding to the
Mo and S sites (isospin) . Right panel: the phase winding under a
three-fold rotation. The green axes indicate the rotation of local
atomic coordinates that leads to the azimuth dissynchronization.}}
\end{figure}
\end{center}
\end{spacing}

The optical selection rule is rooted in the phase winding of the Bloch
states under rotational symmetry, 3-fold rotation ($\hat{C}_{3}$)
in this case. Given the symmetry-adapted linear combination orbitals,
the azimuthal phase associated with the 3-fold rotation at ${\bf K}_{\pm}$
is readily calculable,

\begin{equation}
\hat{C}_{3}|v({\bf K}_{\pm})\rangle=|v({\bf {\bf K}_{\pm}})\rangle,
\end{equation}

\begin{equation}
\hat{C}_{3}|c({\bf K}_{\pm})\rangle=e^{\mp i2\pi/3}|c({\bf K}_{\pm})\rangle
\end{equation}
where $v$ and $c$ correspond to the valence and conduction band
extrema, respectively. Notice, however, the phase winding associated
with the rotation has two distinct contributions. The first comes
from the Bloch phase shift in stepping from one lattice site to the
next, as in the case of gapped graphene.\cite{Xiao07} The second
phase factor arises as a consequence of dissynchronization of the
azimuthal phase (associated with the magnetic quantum number, $l$)
concommitant with rotation of the local atomic coordinates. These
are schematically illustrated in \textcolor{blue}{Fig.2b}.

Now the chiral optical selecvitity of the valleys can be deduced.
The bottom of the conduction bands at the valleys, dominated by the
$l=0$ $d$-states on Mo, bears an overall azimuthal quantum number
$m_{\pm}=\pm1$, at ${\bf K}_{\pm}$. At the top of the valence bands,
$m_{\pm}=0$. Then for an optical transition at ${\bf K}_{\pm}$,
the angular momentum selection rule indicates that $\Delta m_{\pm}=\pm1$,
corresponding to the absorption of left- and right-handed photons.
Therefore, our DFPT results concerning the close neighborhood of the
valleys, which are the most important to the proposed optical valley
polarization, are in fact ensured by the symmetry of the material. 

There is an important distinction, compared to gapped graphene, in
the microscopic origin of chiral optical selection rule; that is,
the selectivity arises directly from the local relative azimuthal
phase of the atomic orbitals, in contrast to the sublattice-dependent
Bloch phase winding in the case of gapped graphene.\cite{Xiao07}
The bonding in the electronic states in MoS$_{2}$ across the gap
is also considerably more complex, exhibiting richer possibilities
of variation owing to the symmetry of local atomic states. Remarkably,
the selectivity is nearly perfect over the entire valleys, and only
changes sign rapidly across valley boundaries (\textcolor{blue}{Fig.2a}).
This is to say, the entire valley ${\bf K}_{+}$ absorbs almost purely
left-handed photons, whereas the entire valley ${\bf K}_{-}$ purely
right-handed. The perfect valley-constrasting CD is very much conducive
to optical polarization of the valleys. 

Now that we have established the valley-selective CD in monolayer
MoS$_2$, it is also interesting to look at the Berry
curvature, ${\bf \Omega}_{n}({\bf k})$, which, if present, has crucial
influence on the transport properties. Of note, Berry curvature enters
into the semiclassical wavepacket dynamics via an anomalous velocity
perpendicular to the applied electric field ($\sim{\bf E}\times{\bf \Omega}_{n}({\bf k})$),
in addition to the usual group velocity of Bloch bands. \cite{Karplus54,Chang95}
The presence of non-vanishing Berry curvature is possible in the non-centrosymmetric
honeycomb lattice. \cite{Xiao07} In \textcolor{blue}{Fig.3}, we plot
the Berry curvature, $\Omega_{n{\bf k},z}=-2\text{Im}\langle\partial u_{n{\bf k}}/\partial k_{x}|\partial u_{n{\bf k}}/\partial k_{y}\rangle$
along the ${\bf K}_{-}-\Gamma-{\bf K}_{+}$ path. Since the system
has time-reversal symmetry and not inversion symmetry, $\Omega_{n}({\bf k})$
is an odd function in ${\bf k}$ with generally non-zero values, as
expected. The charge carriers' anomalous velocity acquires opposite
signs in the two valleys, exactly canceling each other's contribution
to the transverse current. At equilibrium, freestanding monolayer
MoS$_2$ will not exhibit anomalous Hall effect. Notice
that at ${\bf K}_{\pm}$, the Berry curvatures do not have the particle-hole
symmetry. This is a clear indication that the physics of MoS$_{2}$
cannot be fully captured by a minimalistic two-band model, as distinct
from the case of gapped graphene.{\small{} \cite{Xiao07,Xiao11}}{\small \par}

\begin{spacing}{1}
\begin{center}
\begin{figure}[H]
\centering{}\includegraphics[scale=0.42]{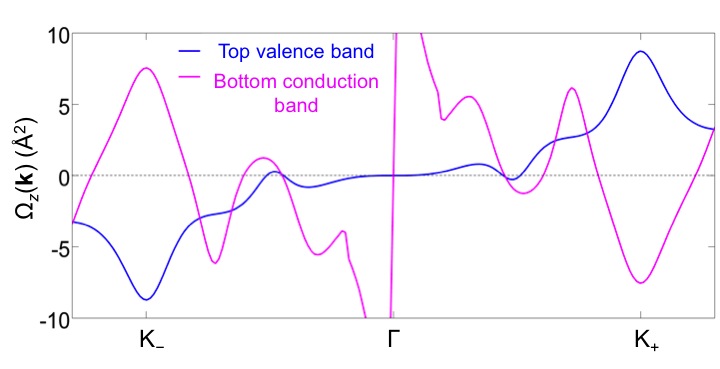}\label{fig:berry}{\small \caption{{\small Berry curvature, $\Omega_{z}({\bf k})$, of bands across the
band gap. The Berry curvature of only the states along the ${\bf K}_{-}-\Gamma-{\bf K}_{+}$
path of the Brillouin zone are plotted. Notice that the value of Berry
curvature is large for the conduction band at the zone center, where
bands are degenerate.}}
}
\end{figure}
\end{center}
\end{spacing}

When valley polarization is induced by, say, valley-selective CD,
only one valley has non-vanishing charge carrier population. This
can then lead to the valley Hall effect.\cite{Yao08} The Berry curvature
across the band edges near ${\bf K}_{\pm}$ is most relevant to photo
excited charge carriers. We see that at the band edges both conduction
and valence bands display significant Berry curvature with opposite
signs. Consequently, when electrons and holes are generated by a circularly
polarized irradiation, both types of charge carriers have an intrinsic
\textit{concurrent} contribution to the Hall conductivity. Our results
indicate that valley Hall effect in this material indeed warrants
experimental assaying. In closing, we also would like to point out
that MoS$_{2}$ is but one of the many transition metal dichalcogenides,
MX$_{2}$, where M = Mo, W, and X = S, Se. They all have identical
crystal structure and are similar electronically. These compounds,
once available in monolayer form, will provide a chemically rich family
of semiconducting atomic membranes for exploring the valley physics,
as we have demonstrated here for MoS$_{2}$.

\bibliographystyle{unsrtnat}
\bibliography{references}

\end{document}